\documentclass[english,twocolumn,showpacs,footinbib]{revtex4-1}

\pdfoutput=1

\usepackage[latin9]{inputenc}
\usepackage{bm}
\usepackage{amsmath}
\usepackage{amssymb}
\usepackage{mathtools}
\usepackage{graphicx}

\makeatletter

\@ifundefined{textcolor}{}
{%
\definecolor{BLACK}{gray}{0}
\definecolor{WHITE}{gray}{1}
\definecolor{RED}{rgb}{1,0,0}
\definecolor{GREEN}{rgb}{0,1,0}
\definecolor{BLUE}{rgb}{0,0,1}
\definecolor{CYAN}{cmyk}{1,0,0,0}
\definecolor{MAGENTA}{cmyk}{0,1,0,0}
\definecolor{YELLOW}{cmyk}{0,0,1,0}
}

\usepackage{hyperref}
\usepackage{xcolor}
\hypersetup{
    colorlinks,
    citecolor=blue,
    filecolor=blue,
    linkcolor=blue,
    urlcolor=blue,
    pdfstartview=FitH
}

\usepackage{amsfonts,amsmath,mathrsfs,epsfig,amsbsy,amssymb,bm,verbatim}
\newcommand{\bra}[1]{\langle {#1} |}
\newcommand{\ket}[1]{| {#1} \rangle}
\newcommand{\expect}[1]{\langle {#1} \rangle}

\newcommand{\ua}{\uparrow}
\newcommand{\da}{\downarrow}
\newcommand{\beq}{\begin{equation}}
\newcommand{\eeq}{\end{equation}}
\newcommand{\beqarray}{\begin{eqnarray}}
\newcommand{\eeqarray}{\end{eqnarray}}
\newcommand{\bal}{\begin{align}}
\newcommand{\eal}{\end{align}}
\newcommand{\eq}[1]{Eq.~(\ref{#1})}
\newcommand{\Ref}[1]{Ref.~\onlinecite{#1}}
\newcommand{\fig}[1]{Fig.~(\ref{#1})}

\allowdisplaybreaks

\makeatother

\usepackage{babel}
\begin{document}

\title{Bound states of a ferromagnetic wire in a
  superconductor} 

\author{Jay D. Sau}
\email{jaydsau@umd.edu}
\affiliation{Condensed Matter Theory Center and Joint Quantum
  Institute, University of Maryland, College Park, Maryland
  20742-4111, USA}
\author{P. M. R. Brydon}
\email{pbrydon@umd.edu}
\affiliation{Condensed Matter Theory Center and Joint Quantum
  Institute, University of Maryland, College Park, Maryland
  20742-4111, USA}

\date{May 11, 2015}
\begin{abstract}
We consider the problem of bound states in strongly
anisotropic ferromagnetic impurities in a
superconductor, motivated by recent experiments that claim to observe
Majorana modes at the 
ends of ferromagnetic wires on a superconducting substrate
[S. Nadj-Perge \emph{et al.}, Science {\bf 346}, 602 (2014)]. 
Generalizing the successful theory of bound states of spherically
symmetric impurities, we consider a wire-like potential using
both analytical and numerical approaches.
We find that away from
the ends of the wire the bound states form bands with pronounced van
Hove singularities, giving rise to subgap peaks in the local density
of states. 
For sufficiently strong magnetization of the wire, we show that this process
generically produces a sharp peak at zero energy in the local density
of states near the ends of the wire. This zero-energy peak has
qualitative similarities to the 
claimed signature of a Majorana mode observed in the aforementioned 
experiment. 
\end{abstract}

\pacs{74.25.Ha,73.20.Hb}


\maketitle

\paragraph{Introduction.}

The antagonistic relationship between superconductivity and ferromagnetism
manifests itself in many diverse ways. For example, it is well known
that a magnetic impurity in a superconductor supports bound
states with energy within the superconducting
gap~\cite{Yu1965,Shiba1968,Rusinov1969}. These 
Yu-Shiba-Rusinov (YSR) states can be directly imaged using scanning
tunneling microscopy (STM)~\cite{Yazdani1997Probing,JiPRL2008}. They
have been  
extensively studied~\cite{BalatskyRMP2006}, and play a key role in 
Kondo-Andreev screening~\cite{FrankeScience2011} and
mediating long-ranged 
antiferromagnetic exchange interactions~\cite{YaoPRL2014}. 
Arrays of magnetic impurities with overlapping YSR states
have recently 
attracted attention as a way to
engineer a topological superconductor with Majorana edge
modes~\cite{ChoyPRB2011,Nadj-Perge2013Proposal,KlinovajaPRL2013,PientkaPRB2013,NakosaiPRB2013,HeimesPRB2014,Brydon2015,Peng2015Strong}. Realizing 
a Majorana mode is not only of fundamental interest, but also of
technological importance, as their 
non-Abelian braiding statistics is key to promising
proposals for a quantum computer~\cite{NayakRMP2008}. 

Much excitement has therefore been generated by recent claims of the
observation of Majorana modes in a ferromagnetic wire on a
superconducting
substrate~\cite{Nadj-Perge2014Observation}. Specifically, STM
measurements reveal a sharply localized peak at zero bias in the local
density of states (LDOS) at the ends of atomically thin iron wires
grown on a lead surface,
which are interpreted as Majorana bound states. Several theoretical
calculations~\cite{Hui2015Majorana,Dumitrescu2015Majorana,LiPRB2014} 
have shown that Majorana modes could occur in this setting,
as had earlier been suggested in mesoscopic 
geometries~\cite{LeeARXIV2009,Chung2011Topological,Duckheim2011Andreev,Takei2012Microscopic,Wang2010Interplay}.

A characteristic aspect of the experimental results is that the
LDOS in the wire shows pronounced subgap
features~\cite{Nadj-Perge2014Observation},  
which resemble the
spectrum of YSR states observed in isolated magnetic
impurities or dimers~\cite{Yazdani1997Probing,JiPRL2008}. This
motivates us to go beyond models of a spherical magnetic
impurity~\cite{Yu1965,Shiba1968,Rusinov1969,BalatskyRMP2006,KimPRL2015},
and consider the ferromagnetic wire itself as a highly anisotropic
scattering center. By focusing upon the physics at subgap energy
scales, we avoid the complexity
of quasi-two-dimensional tight-binding models of the
wire-like impurity~\cite{Nadj-Perge2014Observation,LiPRB2014},
yielding a transparent and intuitive understanding of the 
YSR states in this system. In particular, we wish to
investigate whether or not the features interpreted as Majorana
modes in~\Ref{Nadj-Perge2014Observation} could have a conventional
nontopological origin.  

\begin{figure}

\begin{centering}
\includegraphics[width=1\columnwidth]{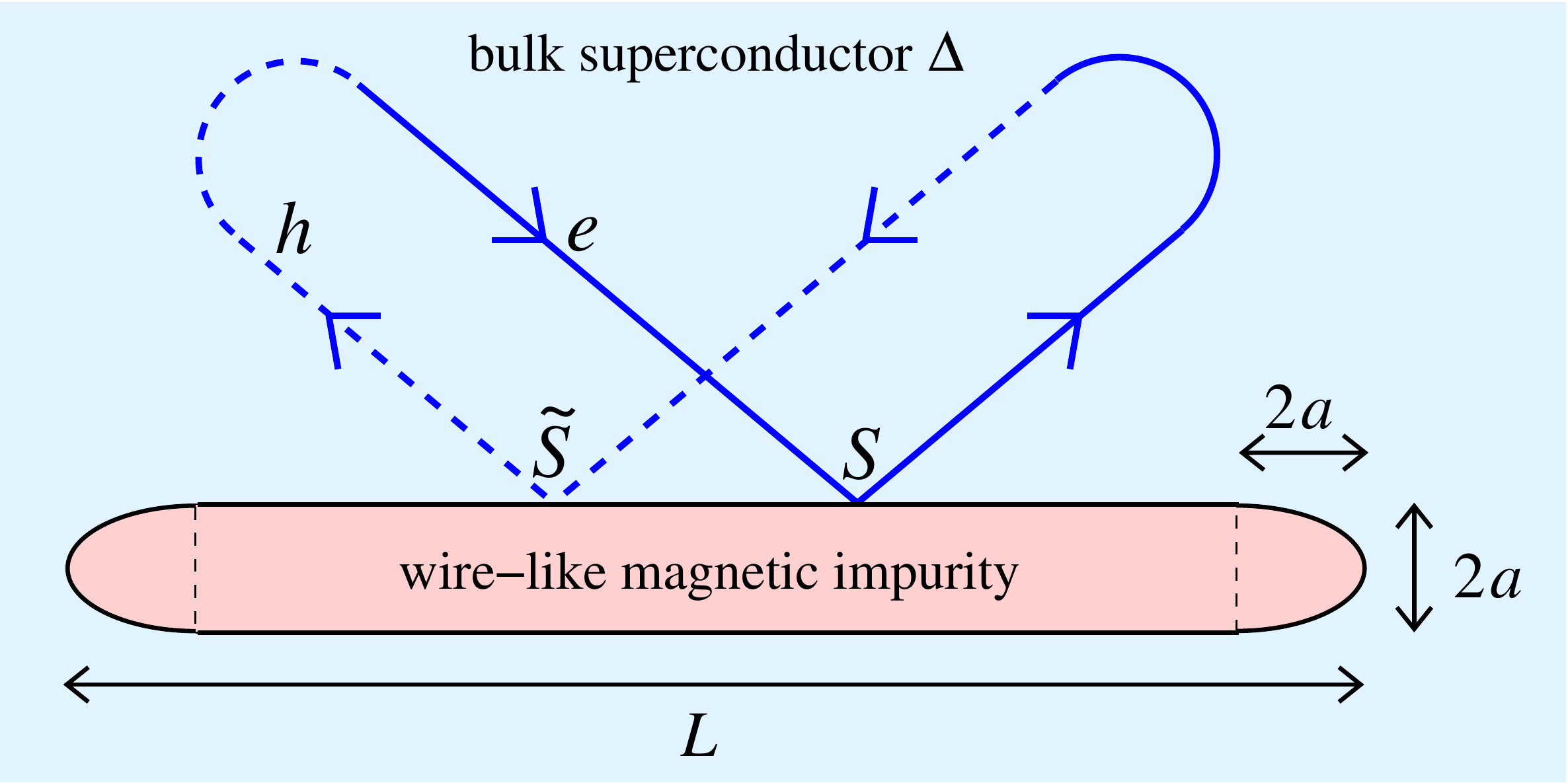}
\par\end{centering}

\caption{(Color online) A wire-like magnetic impurity
  is embedded in a bulk 
  superconductor with gap $\Delta$. The wire has length $L$ and radius
  $a$, with half-ellipsoidal caps at each end. We schematically show a
  YSR state as the scattering of 
  an electron (solid line) at the impurity followed by
  its Andreev reflection as a hole (dashed line), and its subsequent
  scattering and 
  Andreev reflection. Electrons and holes
  undergo conventional reflection at the magnetic impurity, described
  by the scattering matrices $S$ and $\tilde{S}$, respectively. The
  difference in phase for electron- and hole-scattering off the
  impurity is compensated by the energy-dependent phase
  acquired by Andreev reflection in the superconductor.  
\label{fig:schematic}}
\end{figure}

In this letter we present a study of the YSR states in a
ferromagnetic wire-like impurity with smooth ends, shown schematically
in~\fig{fig:schematic}. We start by developing a general
approach for obtaining the energies and wavefunctions of the YSR states
in terms of an eigenvalue problem involving the normal-state scattering
matrices. We apply this formalism to a long wire-like ferromagnetic
impurity and calculate the energy spectra analytically using a
semiclassical approximation. This reveals that the YSR bands in the
wire typically possess van Hove singularities (VHSs) 
which appear as peaks in the LDOS of the wire. As the
wire tapers to its end, these features move towards the gap edges. For
sufficiently strong magnetization of the wire, the VHSs of the two
particle-hole conjugated branches cross 
at zero energy, giving a sharply-localized peak 
in the LDOS near the wire end. We confirm this scenario by 
numerical computation of the scattering matrix of the impurity. The
numerical results for the LDOS 
qualitatively resemble key features of the
experiment~\cite{Nadj-Perge2014Observation}. Finally, we show that
including spin-orbit coupling in the impurity allows the appearance of
Majorana modes inside a small energy gap.  

\paragraph{Scattering matrix formalism for bound states.}

We  first examine a bound state wavefunction
localized at a magnetic impurity in a superconductor, with energy lying
within the gap, i.e. $|E|<\Delta$~\footnote{As suppression
  of the gap near the wire is not
  observed~\cite{Nadj-Perge2014Observation}, we
  follow~\cite{ChoyPRB2011,Nadj-Perge2013Proposal,KlinovajaPRL2013,PientkaPRB2013,NakosaiPRB2013,HeimesPRB2014,Brydon2015,Peng2015Strong,Hui2015Majorana,Dumitrescu2015Majorana,LiPRB2014}
  and 
  assume a constant gap.}. Parametrizing the bound state
energy through $E=\Delta\cos{\varphi}$, away from the impurity the
wavefunction satisfies   
\beq
-\left(\frac{\hbar^2\nabla^2}{2m}+\mu\right)\psi=\Delta(-i\tau_y+\cos{\varphi}\tau_z)\psi\,,
\eeq
where $m$ is the effective mass and $\mu$ the chemical potential in
the bulk superconductor. The solution can be expanded in eigenstates
$(1,e^{\pm i\varphi})^T$ of the matrix 
$(-i\tau_y+\cos{\varphi}\tau_z)$ and angular momentum channels  
\begin{eqnarray}
\psi({\bf r}) &=&\sum_{l,m,\sigma} \left\{h_l^{(-)}(k_- r)a_{l,m,\sigma}\left(\begin{array}{c}1\\e^{i\varphi}\end{array}\right)\right.\nonumber\\
&&\left.+  h_l^{(+)}(k_+ r)b_{l,m,\sigma}\left(\begin{array}{c}1\\e^{-
    i\varphi}\end{array}\right)\right\}Y^m_l(\hat{\bf r})\zeta_\sigma\,,
\end{eqnarray}
where $a_{l,m,\sigma}$ and $b_{l,m,\sigma}$ are the scattering
coefficients, $\zeta_\sigma$ is a spinor, and 
the Hankel functions obey
\beq
-\left(\frac{\hbar^2\nabla^2}{2m}+\mu\right)h_l^{(\pm)}(k_{\pm}r)=\pm i \Delta\sin{\varphi}h_l^{(\pm)}(k_{\pm}r)\,,
\eeq
with $k_{\pm} = [2m(\mu \pm i\Delta\sin\varphi)]^{1/2}/\hbar$.
In the asymptotic limit $|k_{\pm}|r\gg
1$, and with $\mu\gg \Delta$, we approximate  
\begin{eqnarray}
\psi({\bf r})&\approx&\frac{e^{-\frac{r \Delta\sin\varphi}{\hbar
      v_F}}}{k_Fr}\sum_{l,m,\sigma}\left\{  e^{-i[k_Fr - (l+1)\pi/2]}a_{l,m,\sigma}\left(\begin{array}{c}1\\e^{i\varphi}\end{array}\right)\right.\nonumber\\
&&\left.+ e^{i[k_Fr - (l+1)\pi/2]}b_{l,m,\sigma}\left(\begin{array}{c}1\\e^{-
    i\varphi}\end{array}\right)\right\}Y^m_{l}(\hat{\bf r})\zeta_\sigma\,, \label{eq:psiapprox}
\end{eqnarray}
where $v_F = \hbar k_F /m$ is the Fermi velocity and $k_F =
\sqrt{2m\mu}/\hbar$ is the Fermi wavevector.

For $r\ll \xi = \hbar v_F /\Delta$ we may ignore the exponential decay
in~\eq{eq:psiapprox}, and the form of 
$\psi({\bf r})$
in the electron sector is then identical to the wavefunction for
scattering off the impurity in the normal
state. The coefficients obey $b = Sa$,
where we have introduced a matrix-vector notation for the normal-state
scattering matrix $S\equiv
S_{l,m,\sigma;l',m',\sigma'}$, and the vectors $a\equiv a_{l,m,\sigma}$ and $b\equiv
b_{l,m,\sigma}$. Similarly, the hole channel wavefunction
is a normal scattering state for the time-reversed Hamiltonian
$\tilde{H}=\sigma_y H^*\sigma_y$. Therefore, the vectors $a$, $b$ also
satisfy $e^{-2i\varphi}b=\tilde{S}a$,
where the time-reversed scattering matrix $\tilde{S}$ is given by
$\tilde{S}=\Lambda \sigma_y S^T\sigma_y \Lambda^\dagger$ and $\Lambda$
is the matrix that time-reverses the orbital indices. 
We hence obtain $a$ and the phase $\varphi$, which determines the
bound state energy $E=\Delta\cos{\varphi}$ by solving the
 eigenvalue problem
\beq
e^{-2i\varphi}a=S^\dagger\tilde{S}a\,. \label{eq:a}
\eeq
The other scattering amplitude $b$ is obtained from $a$ using $b=Sa$.
Equation~(\ref{eq:a}) generalizes the theory of YSR bound
states~\cite{Yu1965,Shiba1968,Rusinov1969} to 
arbitrary geometries.  


\paragraph{Semiclassical theory for a ferromagnetic wire.} As an
application of~\eq{eq:a}, we turn our
attention to a wire-like impurity, which we set to lie along the
$z$-axis. The scattering matrix can be
estimated semiclassically, assuming that the wire is smooth on the
scale of $k_F^{-1}$. This is satisfied for the experiment
reported in~\cite{Nadj-Perge2014Observation}, where in the lead
substrate we have $k_F^{-1}\approx 0.13a_{\text{Pb}}$, with
$a_{\text{Pb}}$ the lattice constant~\cite{AshcroftMermin}.  
The small value of $k_F^{-1}$ makes it reasonable to assume
that the scattering potential changes sufficiently slowly along the
wire such that the parallel component of quasiparticle 
momentum $k_z$ is approximately conserved. For a cylindrically
symmetric wire, the calculation of
the bound-state energies for fixed $k_z$ reduces to the
bound-state problem for an isotropic impurity in two
dimensions. Setting the magnetic moment along the
$z$-axis, the scattering matrix in azimuthal angular momentum
channel $m$ is
\beq
S_m(k_z,z) = e^{i\phi_{m}(k_z,z)
}e^{i b_m(k_z,z)\sigma_z}\,, \label{eq:Sm}
\eeq  
where $\phi_{m}(k_z,z)$ [$\sigma b_{m}(k_z,z)$] is the
  spin-independent (spin-dependent) component of the phase shift
acquired by a scattered spin-$\sigma$ electron, and $\sigma$ is
positive (negative) according as the spin is aligned (antialigned)
with the impurity magnetic moment. 
In the supplemental material we show that
$\phi_m$ and $b_m$ satisfy~\footnote{See the Supplemental 
  Material.}
\begin{multline}
\tan{(\phi_m + \sigma b_m)}\\=\frac{|k_\sigma| J_m(k_\perp
  a)\tilde{J}^\prime_m(k_\sigma 
    a)-k_\perp J^\prime_m(k_\perp a)\tilde{J}_m(k_\sigma a)}{|k_{\sigma}|Y_m(k_\perp
  a)\tilde{J}^\prime_m(k_\sigma a)- k_\perp Y^\prime_m(k_\perp
  a)\tilde{J}_m(k_\sigma a)}, \label{eq:2Dscattprob}
\end{multline}
where $k_\perp^2 = k_F^2 -
k_z^2$, $k_\sigma^2=k_\perp^2- 
k_F^2V_\sigma/\mu$ with $V_\sigma$ the scattering potential
experienced by spin-$\sigma$ particles, $J_m(x)$ and $Y_m(x)$ are
Bessel functions, and $\tilde{J}_m$ is $J_m(x)$ or the modified Bessel
function $I_m(|x|)$ according as $k_\sigma^2$ is positive or negative,
respectively. 
The radius $a$ of the
wire  vanishes towards the ends, giving the
$z$-dependence in~\eq{eq:Sm}.
Reflection and spin-rotation invariance imply
$\phi_m(-k_z,-z)=\phi_m(k_z,z)$
and $b_m(-k_z,-z)=b_m(k_z,z)$. The time-reversed scattering
matrix is  
\beq
\tilde{S}_m(k_z,z)=e^{i \phi_{m}(k_z,z) }e^{-i b_m(k_z,z)
  \sigma_z}\,. \label{eq:TRSm}
\eeq  
The eigenstates of $S_m^\dagger \tilde{S}_m =
\exp(-2ib_m\sigma_z)$ yield the
energy of the spin-$\sigma$ state in the $m$ channel
\beq
E_{m,\sigma}(k_z,z)=\sigma\Delta\cos{[b_m(k_z,z)]}\,. \label{eq:YSRwire}
\eeq

\begin{figure}
\includegraphics[width=\columnwidth]{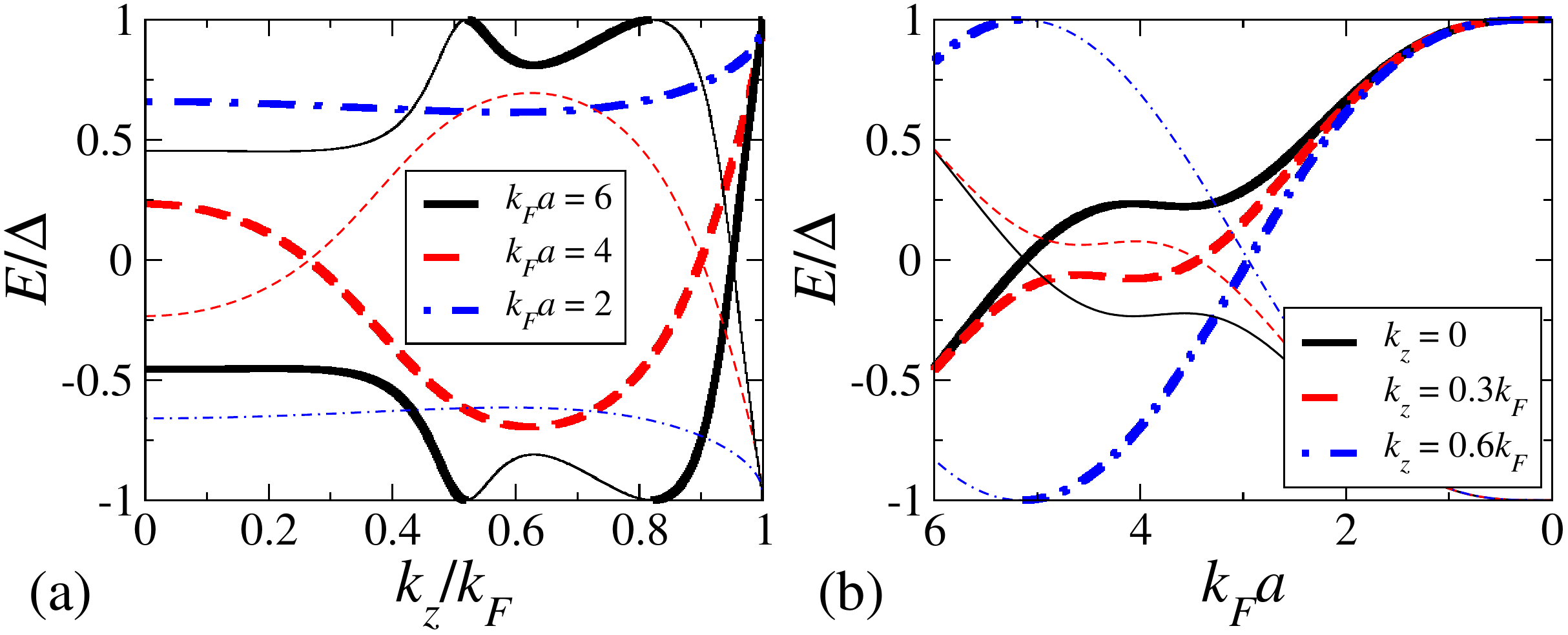}
\caption{(Color online) Semiclassical results for the $m=0$ YSR bands
  in a magnetic wire. (a) Dispersion of the bands for different
  wire thicknesses $a$. (b) Dependence of the bound state energy on the
  wire thickness $a$ at various fixed $k_z$. The spin-$\uparrow$
  ($\downarrow$) branch is represented by the thick (thin) lines. We set
  the potentials 
  $V_\uparrow = 0.04\mu$ and $V_\downarrow = 0.6\mu$. 
\label{fig:semiclassics}}
\end{figure}

\begin{figure}

\begin{centering}
\includegraphics[width=0.465\columnwidth]{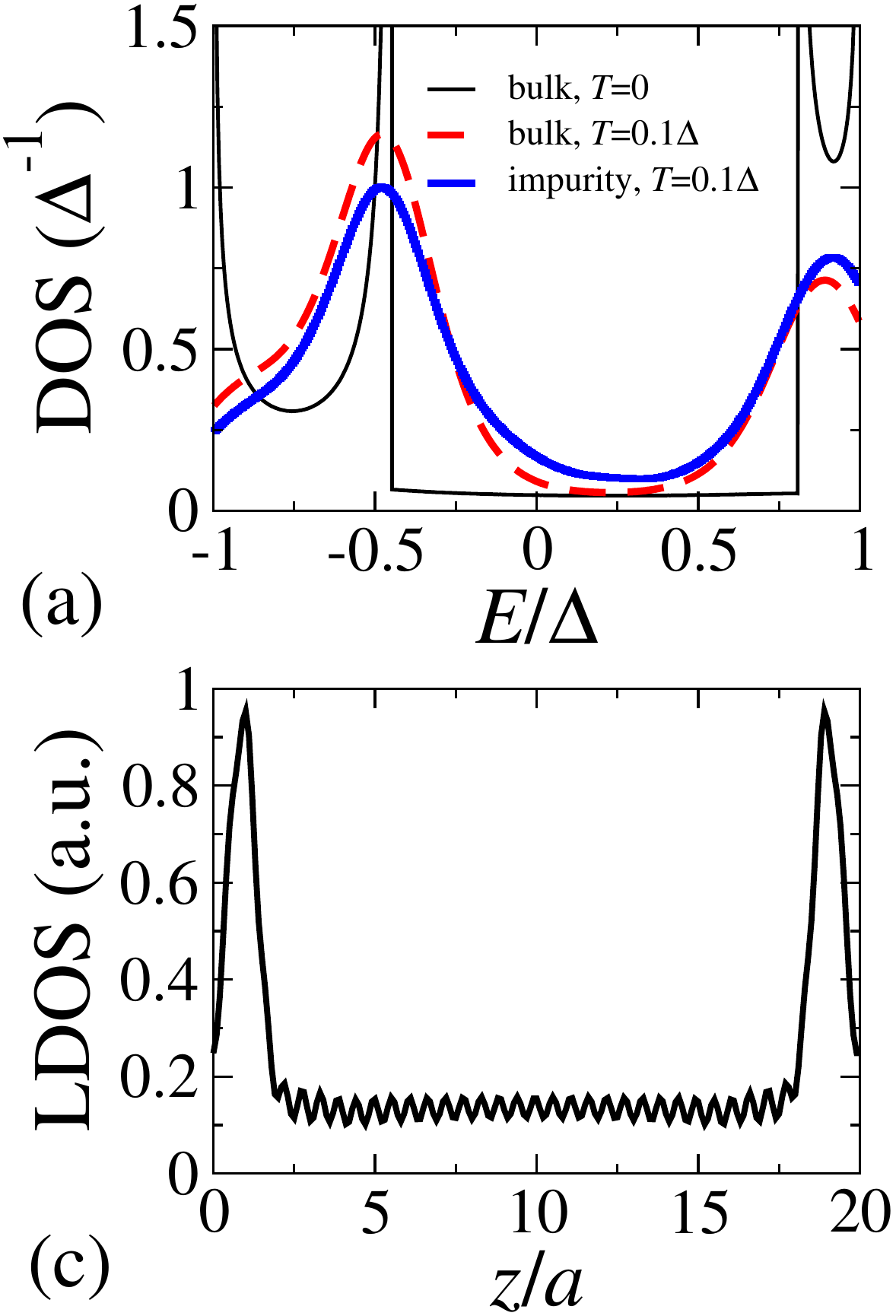}
\hspace*{0.005\columnwidth}\includegraphics[width=0.515\columnwidth]{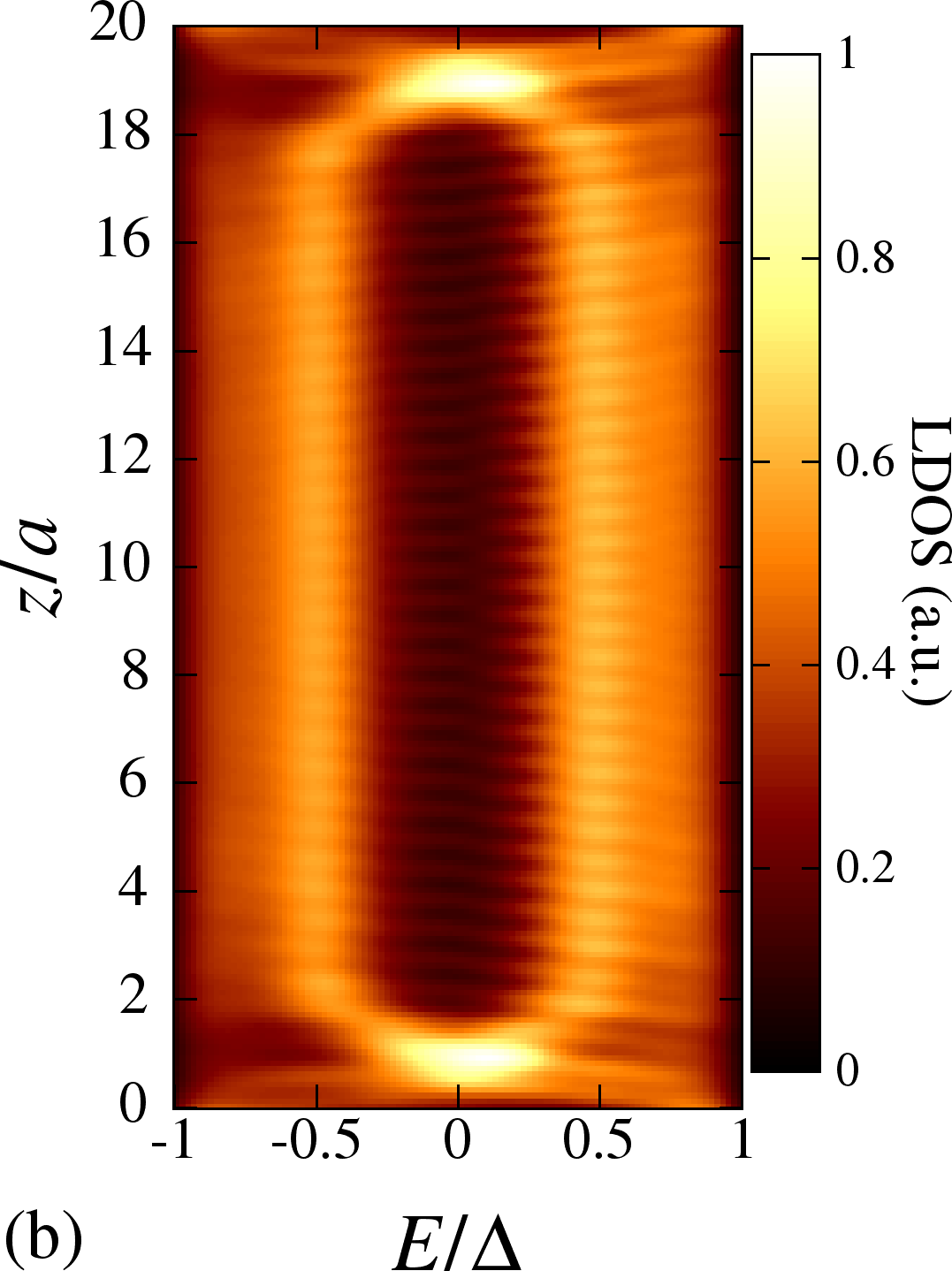}
\par\end{centering}

\caption{(Color online) Numerical results for the
  $m=0$ YSR states at the
  wire-like impurity. (a) Comparison of the total
    spin-$\uparrow$ subgap DOS for the infinite wire
    (``bulk'') and the finite impurity. (b) Energy-resolved LDOS as a
  function of position $z$  
  along the wire, with maximum value normalized to one. (c)
  Zero-energy LDOS as a function of position along the 
  wire. We take the same potential for the wire as
  in~\fig{fig:semiclassics}. The wire length is chosen to be
  $L=20a$, and the Fermi wavevector is related to the radius by $k_Fa
  = 6$. We restrict the expansion of the $T$ matrix to angular
  momentum channels with $l \leq 70$. 
\label{fig:numerics}}
\end{figure}

We consider a magnetic potential
sufficient to cause the $m=0$ YSR bands to
cross zero energy. Typical results for the $m=0$ bands
are shown in~\fig{fig:semiclassics}; results for $m>0$ are
  provided in the supplemental material~\cite{Note2}. The YSR bands
  generically have VHSs, which produce 
strong peaks in the DOS~\cite{Peng2015Strong} as shown
in~\fig{fig:numerics}(a). In particular, the peak at $\approx
-0.5\Delta$ corresponds to the $k_z=0$ 
VHS, while the singularity at $\approx
0.8\Delta$ matches the feature centered about $k_z/k_F \approx 0.7$. 
These VHSs are quite robust to changes in the scattering
potentials~\cite{Note2}. 
The reduction of the radius at the ends of the wire
weakens the effective strength of the impurity, causing
the bands to shift towards the gap edge, as shown
in~\fig{fig:semiclassics}(b). The VHSs will therefore
also shift, and so the LDOS near the wire ends 
is qualitatively different from that in the middle. We estimate the
length scale of this reconstruction by 
modeling the YSR band near the VHS with the effective
Hamiltonian $H = -\hbar^2\partial_z^2/2\tilde{m} + (z-z_0)V_0$, where
the linear potential accounts for the movement of the VHS,
which crosses zero energy at $z=z_0$.
The zero-energy solution is the Airy function
$\mbox{Ai}([z-z_0]/l_{\text{eff}})$, with characteristic length scale
$l_{\text{eff}} = (\hbar^2/2\tilde{m}V_0)^{1/3}$. Estimating the
effective mass $\tilde{m}\sim k_F^2/\Delta$ and the rate of change in the
position of the VHS $V_0\sim \Delta k_F$, we 
deduce that the peak in the LDOS will be localized over a region
comparable to $l_{\text{eff}}\sim k_F^{-1}$. The peak at zero 
energy is expected to be particularly pronounced, due to the
constructive effect of the crossing of the 
VHSs of the spin-$\ua$ and -$\da$ branches.

\paragraph{Numerical results:} Here we confirm the intuitive
semiclassical picture
of the YSR physics by numerical computation of the
scattering matrix, and 
furthermore calculate the  LDOS along the wire. 
Our starting point is the relationship between the scattering matrix
$S$ and the $T$ matrix~\cite{Merzbacher} 
\beq
S_{l,l';m,\sigma}=\delta_{l,l'}-4\pi i k_F \expect{l,m|T_\sigma|l',m}\,,
\eeq 
where $\expect{{\bf r}|l,m}=j_l(k_F r)Y_{l}^m(\hat{\bf
  r})/\sqrt{2\pi}$. 
The $T$-matrix for spin-$\sigma$ scattering is formally given by
$T_\sigma = V_\sigma[1 - G_0V_\sigma]^{-1}$, where 
$G_{0}({\bf r},{\bf r}') = -e^{ik_F|{\bf r}-{\bf r'}|}/4\pi|{\bf
  r}-{\bf r'}|$ is the free Green's function. 
The $T$ matrix is evaluated by expanding the Green's
function and potential in terms of angular momentum
eigenstates on a real-space grid inside the wire. Having calculated
the scattering matrix, we then use~\eq{eq:a} to obtain
the energy levels, from which we extract the total
DOS of the wire plotted in~\fig{fig:numerics}(a). There is
excellent agreement with the thermally-broadened bulk semiclassical
results.

To obtain the \emph{local} DOS we compute the
wavefunctions within the 
impurity. In the $T$-matrix formalism, the scattering wavefunction
$|\psi\rangle$ is related to the bare wavefunction $\ket{\phi}$ by
$\ket{\psi}=(1-G_0 V)^{-1}\ket{\phi}$.  We obtain the
bound states by choosing $\phi_{n;m,\sigma}({\bf r}) =
\sum_{l}a_{n;l,m,\sigma}\expect{{\bf r}|l,m}$, where $a_{n}$ is the $n$th
eigenvector of~\eq{eq:a} corresponding to a state with energy
$E_{n;m,\sigma}=\sigma E_{n;m}$. The particle- and
hole-components of the spin-$\uparrow$ bound state 
wavefunctions are then 
\begin{subequations}
\begin{align}
& u_{n;m,\uparrow}({\bf r})=\sum_{l}\langle {\bf r} |(1-G_0
    V_\uparrow)^{-1}\ket{l,m}a_{n;l,m,\uparrow}\,,\\  
& v_{n;m,\uparrow}({\bf r})=\sum_{l}\bra{{\bf r}}(1-G_0 V_\downarrow)^{-1}\ket{l,m}a_{n;l,m,\uparrow}\,.
\end{align}
\end{subequations}
Summing over the bound states and integrating over the
direction perpendicular to the wire axis, we obtain the LDOS as a
function of the position along the wire
\beqarray
\rho_{m}(z,E) &\propto& \int_{r_\perp<a(z)} d^2r_\perp \sum_{n}\left\{|u_{n;m,\uparrow}({\bf r})|^2\delta(E - E_{n;m})
\right.\notag \\
&& \left.+ |v_{n;m,\uparrow}({\bf r})|^2\delta(E + E_{n;m})\right\}\sqrt{E^2-E_{n;m}^2}\,.
\eeqarray
The factor of
$\sqrt{E^2-E^2_{n;m}}$ is due to the greater 
delocalization of the bound state near the gap edges.
In evaluating the LDOS, we include thermal broadening of the
$\delta$-functions at a temperature  $k_BT = 0.1\Delta$. 

We plot the LDOS in~\fig{fig:numerics}(b). In the bulk of the wire
(i.e. $2a<z<18a$) the LDOS shows very weak $z$-dependence. The main
features are broad peaks centered at $|E|\approx0.5\Delta$ and
$0.8\Delta$, which correspond to the VHS of the
wire-limit bands. Within the end caps (i.e. $z<2a$,
$z>18a$), however, there is substantial modification of the LDOS: The
peaks at $\pm 0.5\Delta$ move towards $\mp \Delta$, 
and their crossing at zero energy gives a strong peak in the LDOS, as
anticipated above. The
zero-energy LDOS along the wire is shown in~\fig{fig:numerics}(c),
and reveals that this peak is sharply localized at the 
ends. The LDOS for the $m=1,2$ channels show similar
behaviour. For $m>2$, however, the potential is not
strong enough to cause the bulk YSR bands to cross zero energy. As
such, the VHSs of these bands do not shift through zero energy near the
end of the wire, and they do not contribute to
the enhancement of the zero-energy LDOS there~\cite{Note2}. 

Our model results show strong qualitative
similarities to the experimental data 
in~\Ref{Nadj-Perge2014Observation}. Specifically, we 
reproduce the subgap peaks in the bulk of the wire and the sharply
localized zero-bias peaks at the wire ends. In our model, however, the
latter feature does not arise from a bound state (Majorana or
otherwise) distinct from the bulk YSR bands, but rather originates
from the shifting of the bulk bands at the ends of the wire. The 
resulting zero-energy peak in the LDOS at the wire end is rather generic,
requiring only that the impurity potential be strong enough that the
VHSs of the YSR bands cross zero energy as the
radius of the wire vanishes. 
A quantitative comparison with~\cite{Nadj-Perge2014Observation} 
would require us to determine the relative contributions of each $m$
to the STM measurements, which
is well beyond the scope of the current work. Nevertheless, our 
considerations show
that the observed zero-energy LDOS features at the wire 
ends cannot be regarded as an
unambiguous signature of a Majorana mode.

\paragraph{Majorana modes in spin-orbit coupled wires:}
So far we have restricted our attention to purely ferromagnetic
wire-like impurities. Realizing a topologically
non-trivial state in such a wire, however, requires
SOC either within the wire~\cite{SauPRB2010} or the
superconductor~\cite{LeeARXIV2009,Duckheim2011Andreev}.  Here we 
modify our semiclassical formalism 
to account for the former scenario.
Neglecting the $z$-dependence of the wire profile, we
generalize the scattering matrix~\eq{eq:Sm} and its
time-reversed form~\eq{eq:TRSm} as $S_m(k_z) = \exp(i[\phi_m(k_z) +
  {\bf b}_m(k_z)\cdot\bm\sigma])$ and $\tilde{S}_m(k_z) = \exp(i[\phi_m(-k_z) -
  {\bf b}_m(-k_z)\cdot\bm\sigma])$, respectively,
where the orientation of ${\bf b}_{m}(k_z)$ depends on $k_z$
due to  the SOC in the impurity.
For a mirror-symmetric geometry, we choose 
$\phi_m(k_z)=\phi_m(-k_z)$ and ${\bf b}_m(k_z)=b_m(k_z)\sin{\theta_m(k_z)}{\bf x} +
b_m(k_z)\cos{\theta_m(k_z)}{\bf z}$ where  $b_m(k_z)=b_m(-k_z)$ and
$\theta_m(k_z) = -\theta_m(-k_z)$ is the angle of the momentum-dependent
magnetic moment with the wire axis. Solving~\eq{eq:a}, we obtain the
YSR bands 
\beq
\widetilde{E}_{m,\pm}(k_z)=\pm\Delta\sqrt{1-\cos^2{\theta_m(k_z)}\sin^2{b_m(k_z)}}\,,
\eeq
showing that the SOC gaps
the spectrum~\eq{eq:YSRwire} of the wire. 

The wire with SOC is in symmetry class $D$, which is
characterized by a $\mathbb{Z}_2$ topological number, specifically
a Pfaffian. In our semiclassical model this is determined by
the spectrum at $k_z= 0$. Since the SOC is
irrelevant here, the Pfaffian
$W=\text{sgn}\{E_{0,\uparrow}(k_z=0)\}$, where $E_{m,\sigma}(k_z)$ is
the dispersion in the 
absence of SOC~[\eq{eq:YSRwire}]. The topologically
non-trivial state corresponds to $W=-1$. 
From~\fig{fig:semiclassics} we see that for the parameters chosen
in our numerical study, adding SOC places the
wire in the topological regime, allowing it to support Majorana modes at its
ends. These zero-energy Majorana modes would appear in the background
of the low-energy LDOS in~\fig{fig:numerics}(b). At finite
temperatures, however, all these subgap features will broaden
considerably, making experimental interpretation
difficult~\cite{Dumitrescu2015Majorana}.   

\paragraph{Conclusions.} 

In this letter we have studied the YSR states in a
wire-like ferromagnetic impurity embedded in a
superconductor. Such systems are currently of great interest as a
possible platform for realizing Majorana
modes~\cite{Nadj-Perge2014Observation,Hui2015Majorana,Dumitrescu2015Majorana,LiPRB2014}. Using
both analytical and numerical methods, we find that
the YSR bands typically display VHSs,
which appear as pronounced peaks in the wire LDOS, and which move
towards the gap edges at the ends of the wire.
For sufficiently strong magnetic scattering, the two
YSR branches cross near the wire ends, giving rise to sharply-localized
features in the LDOS at zero energy. Addition of SOC in the wire
can stabilize a topological regime, but any Majorana features would
appear in the background of an already elevated LDOS near the wire
ends. The localized zero-energy peak in the LDOS obtained here is
qualitatively consistent with recent STM 
measurements~\cite{Nadj-Perge2014Observation} and is independent of
the existence of Majorana modes. A conclusive
demonstration of Majorana modes will need to distinguish them from the
zero-energy conductance feature presented here.

\begin{acknowledgments}
We thank A. Akhmerov, S. Das Sarma, H.-Y. Hui, A. M. Lobos,
R. M. Lutchyn, and D. Rainis for useful discussions. This work is
supported by JQI-NSF-PFC and Microsoft Q. 
\end{acknowledgments}

\bibliographystyle{apsrev4-1}
\bibliography{RenormLen}

\onecolumngrid


\newpage
\widetext

\renewcommand{\theequation}{S\arabic{equation}}
\renewcommand{\thefigure}{S\arabic{figure}}
\setcounter{page}{1}
\setcounter{equation}{0}
\setcounter{figure}{0}

\begin{center}
\textbf{{\large Supplemental Material for\\[0.5ex]
Bound states of a ferromagnetic wire in a superconductor}}\\[1.5ex]
Jay D. Sau and P. M. R. Brydon
\end{center}

\section{Semiclassical theory of the infinite ferromagnetic wire}

\subsection{Scattering phase shifts}

We consider an infinite ferromagnetic wire impurity of radius
$a$, lying along the $z$ axis. The Hamiltonian 
for spin-$\sigma$ particles is written
\beq
H = -\frac{\hbar^2\pmb{\nabla}^2}{2m} -
\mu + V_\sigma\Theta(a - \rho) \label{seq:Hwire}
\eeq
where $\mu = \hbar^2 k_F^2/2m$ is the chemical potential and
$V_{\sigma}$ is the impurity 
potential of the wire. Due to rotational symmetry about the $z$-axis
and translational symmetry along the wire, solutions of~\eq{seq:Hwire}
have the form
\beq
\psi_\sigma({\bf r}) = \psi_{m,k_z,\sigma}(\rho)e^{im\phi}e^{ik_zz}\,,
\eeq
where $m$ is the azimuthal quantum number and $k_z$ is the momentum
along the $z$ axis. A general ansatz for the radial function
$\psi_{m,k_z,\sigma}(\rho)$ is
\beqarray
\psi_{m,k_z,\sigma}(\rho) & = & \Theta(\rho - a)\left\{\alpha_{m,k_z,\sigma} J_m(k_\perp\rho) +
\beta_{m,k_z,\sigma}Y_m(k_\perp\rho)\right\} \notag \\
&& + \Theta(a-\rho)\{\Theta(\hbar^2k_\perp^2/2m - V_\sigma)c_{m,k_z,\sigma}
J_m(|k_\sigma|\rho) + \Theta(-\hbar^2k_\perp^2/2m + V_\sigma)
c_{m,k_z,\sigma}I_m(|k_\sigma|\rho)\}\,.
\eeqarray
Here $J_m(x)$ and $Y_m(x)$ are Bessel functions of the first and
second kinds, respectively, while $I_m(x)$ is the
modified Bessel function of the first kind. The wavevector outside of
the wire is $k^2_\perp = k_F^2-k_z^2$, while $k^2_\sigma
= k_\perp^2 - 2m V_\sigma/\hbar^2$ is the 
wavevector or decay length within the cylinder, according as the term
inside the absolute value sign is positive or negative,
respectively. The coefficients $\alpha_{m,k_z,\sigma}$,
$\beta_{m,k_z,\sigma}$, and $c_{m,k_z,\sigma}$ are $C$ numbers.

Far away from the scattering center (i.e. $\rho \gg a$) we have the
asymptotic form 
\beqarray
\psi_{m,k_z,\sigma}(\rho) &\sim& \alpha_{m,k_z,\sigma} \frac{1}{\sqrt{k_\perp\rho}}\cos(k_\perp\rho - m\pi/2 -
\pi/4) + \beta_{m,k_z,\sigma}\frac{1}{\sqrt{k_\perp\rho}}\sin(k_\perp\rho - m\pi/2 -
\pi/4) \notag \\
& \propto & \frac{1}{\sqrt{k_\perp\rho}}\cos(k_\perp\rho - m\pi/2 - \pi/4 + \delta_{m,k_z,\sigma})
\eeqarray
where $\delta_{m,k_z,\sigma} = \phi_m(k_z) + \sigma b_m(k_z) = -\arctan(\beta_{m,k_z,\sigma}/\alpha_{m,k_z,\sigma})$ is the
scattering phase shift. The ratio
$\beta_{m,k_z,\sigma}/\alpha_{m,k_z,\sigma}$ is determined from the
boundary conditions at the edge of the wire
\beqarray
\psi_{m,k_z,\sigma}(\rho = a^{+}) &=& \psi_{m,k_z,\sigma}(\rho =
a^{-})\,, \\ 
\partial_\rho\psi_{m,k_z,\sigma}(\rho = a^{+}) &=&
\partial_\rho\psi_{m,k_z,\sigma}(\rho = a^{-})\,. 
\eeqarray
After straightforward algebra, we deduce 
\beq
\tan(\delta_{m,k_z,\sigma}) = \tan(\phi_m(k_z) + \sigma b_{m}(k_z))
 = \begin{dcases}
\frac{k_\sigma J_n(ka)J^\prime_n(k_\sigma a) - k J^\prime_n(ka)J_n(k_\sigma a)}{k_\sigma
  Y_n(ka)J^\prime_n(k_\sigma a) - k
  Y^\prime_n(ka)J_n(k_\sigma a)}\,, & \hbar^2k_\perp^2/2m > V_\sigma \\
\frac{|k_\sigma| J_n(ka)I^\prime_n(|k_\sigma| a) - k J^\prime_n(ka)I_n(|k_\sigma| a)}{|k_\sigma|
  Y_n(ka)I^\prime_n(|k_\sigma| a) - k
  Y^\prime_n(ka)I_n(|k_\sigma| a)}\,, & \hbar^2k_\perp^2/2m < V_\sigma
\end{dcases} \label{eq:phaseshift}
\eeq
This is given by Eq.~(7) in the main text.

\subsection{YSR bands}

\begin{figure}

\begin{centering}
\includegraphics[width=0.75\columnwidth]{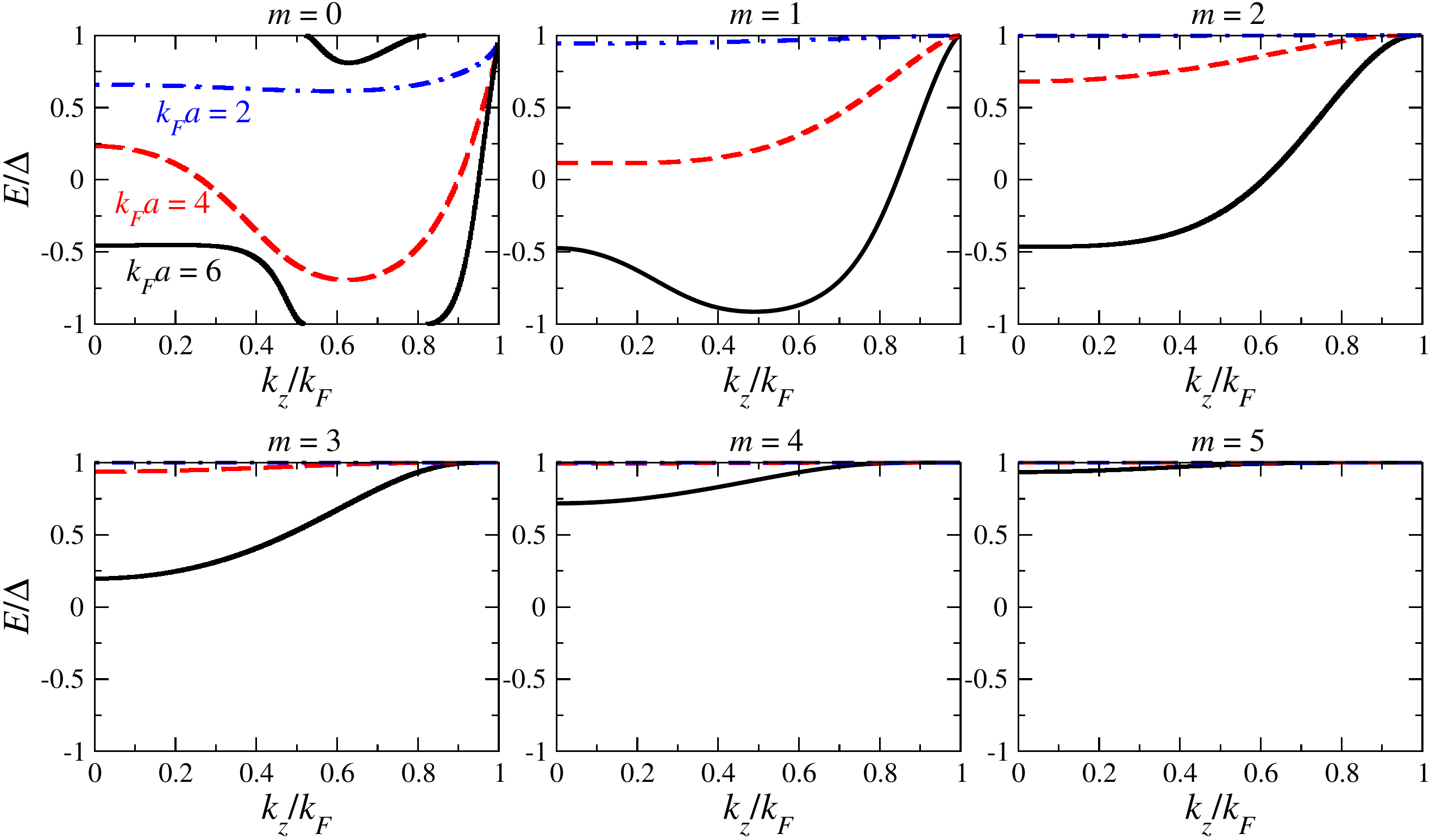}
\par\end{centering}

\caption{Semiclassical results for the dispersion of $0\leq m \leq 5$
  spin-$\uparrow$ YSR bands in a magnetic wire for different wire
  thicknesses $a$. We set
  the potentials 
  $V_\uparrow = 0.04\mu$ and $V_\downarrow = 0.6\mu$. The $m=0$
  results are given as Fig. (2)(a) of the main text.
\label{fig:YSRbands}}
\end{figure}

The YSR bands are obtained from the formula [Eq.~(9) of main text]
\beq
E_{m,\sigma}(k_z,z) = \sigma \Delta \cos[b_{m}(k_z,z)]
\eeq
where $b_{m}(k_z,z)$ is calculated from~\eq{eq:phaseshift}. We plot
the $0\leq m\leq 5$ spin-$\uparrow$ bands as a function of $k_z$ for
different wire 
radius $a$ in~\fig{fig:YSRbands}. We note that all bands show a
pronounced van Hove singularity (VHS) at $k_z=0$. The impurity
potentials are such that the $m\leq2$ bands close the gap in the bulk
of the wire, i.e. they cross zero energy, with the VHS located at
negative energies. As the width of the wire is reduced, these channels
are therefore expected to contribute to the enhancement of the local
density of states (LDOS) at
zero 
energy near the end of the wire. On the other hand, the impurity is
not strong enough to cause the $m>2$ YSR bands to cross zero energy,
and these channels therefore do not contribute to the enhancement of
the zero-energy LDOS at the ends of the wire.

\begin{figure}

\begin{centering}
\includegraphics[width=0.65\columnwidth]{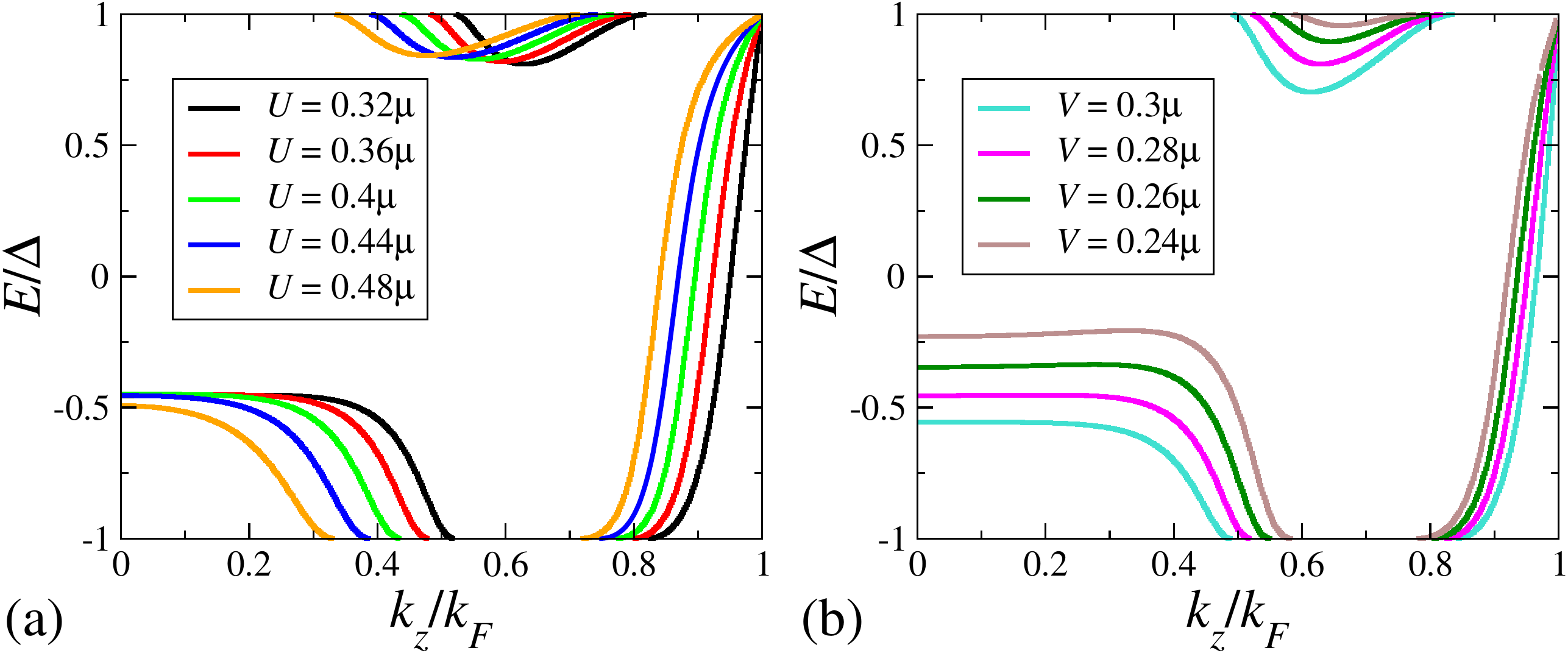}
\par\end{centering}

\caption{Semiclassical results for the dispersion of the $m=0$ 
  spin-$\uparrow$ YSR band upon varying (a) the nonmagnetic potential
  $U$ and (b) the magnetic potential $V$. We set $k_Fa = 6$ in all
  figures. In (a) we take $V=0.28\mu$ while in (b) we have $U=0.32\mu$.
\label{fig:UV}}
\end{figure}

The qualitative form of the YSR bands is robust to significant changes
in the nonmagnetic component of the potential
$U=\tfrac{1}{2}(V_\uparrow + V_\downarrow)$ as shown
in~\fig{fig:UV}(a): increasing $U$ by 50\% produces only minor changes
in the location of the van Hove singularities. The
bands are more sensitive to changes in the magnetic potential
$V=\tfrac{1}{2}(V_\downarrow-V_\uparrow)$ as seen in~\fig{fig:UV}(b),
but still there is a large range of $V$ values for which the
important van Hove singularity at $k_z=0$ is present. This indicates
the generic nature of our results and the absence of fine-tuning.

\section{Numerical results for wire LDOS}

\begin{figure}

\begin{centering}
\includegraphics[height=4.7cm]{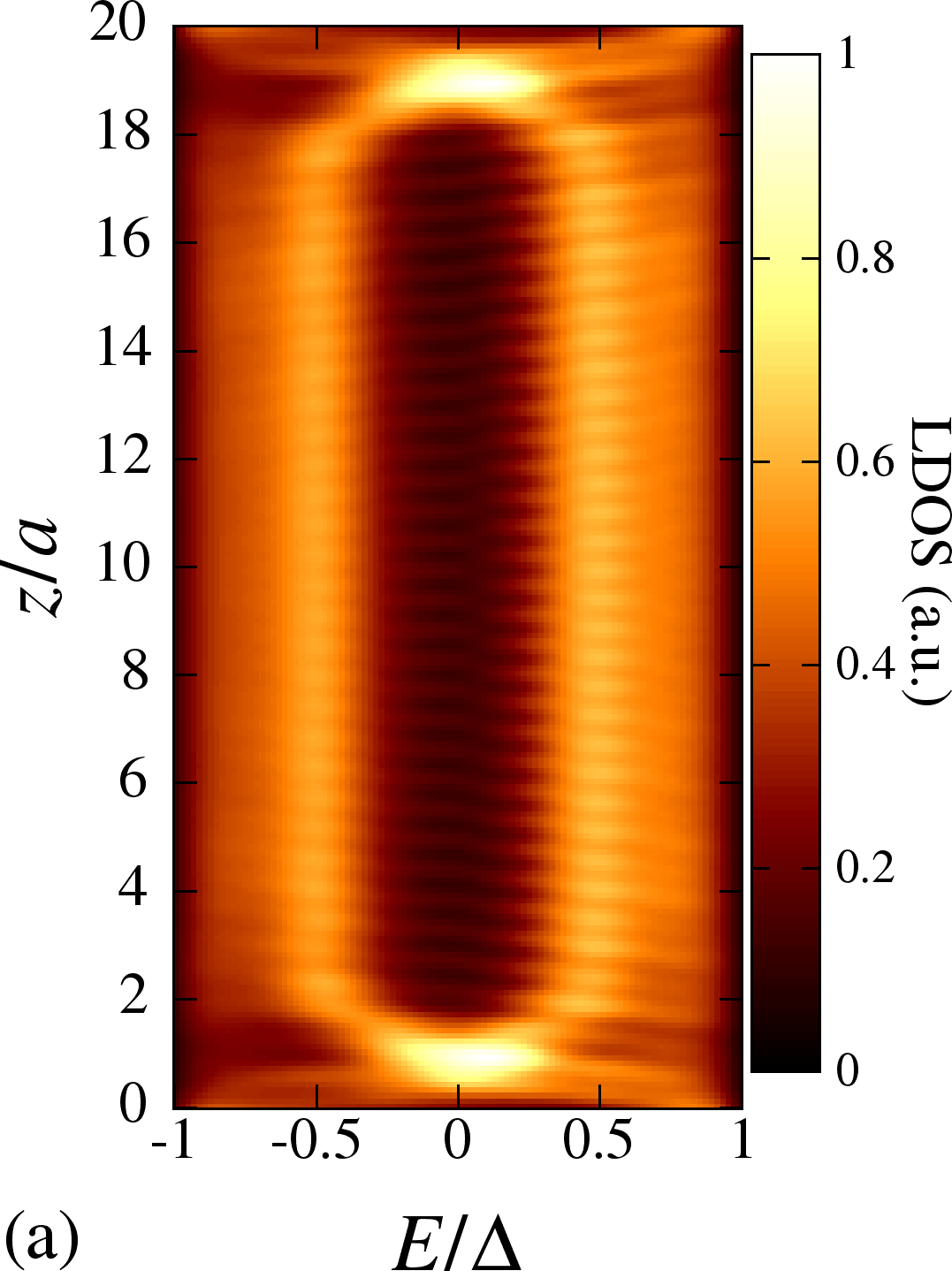}\hspace{0.02\columnwidth}\includegraphics[height=4.7cm]{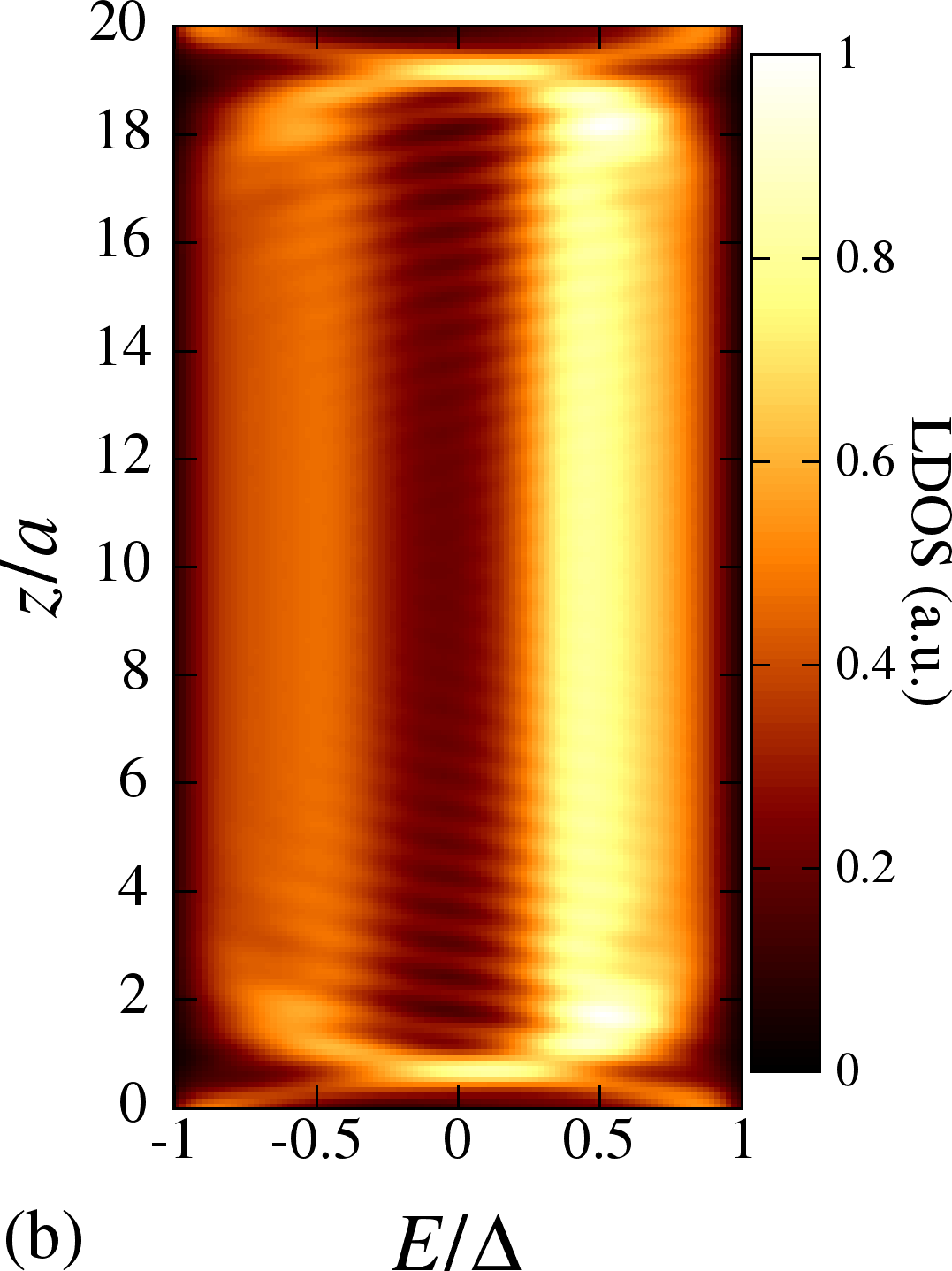}\hspace{0.02\columnwidth}\includegraphics[height=4.7cm]{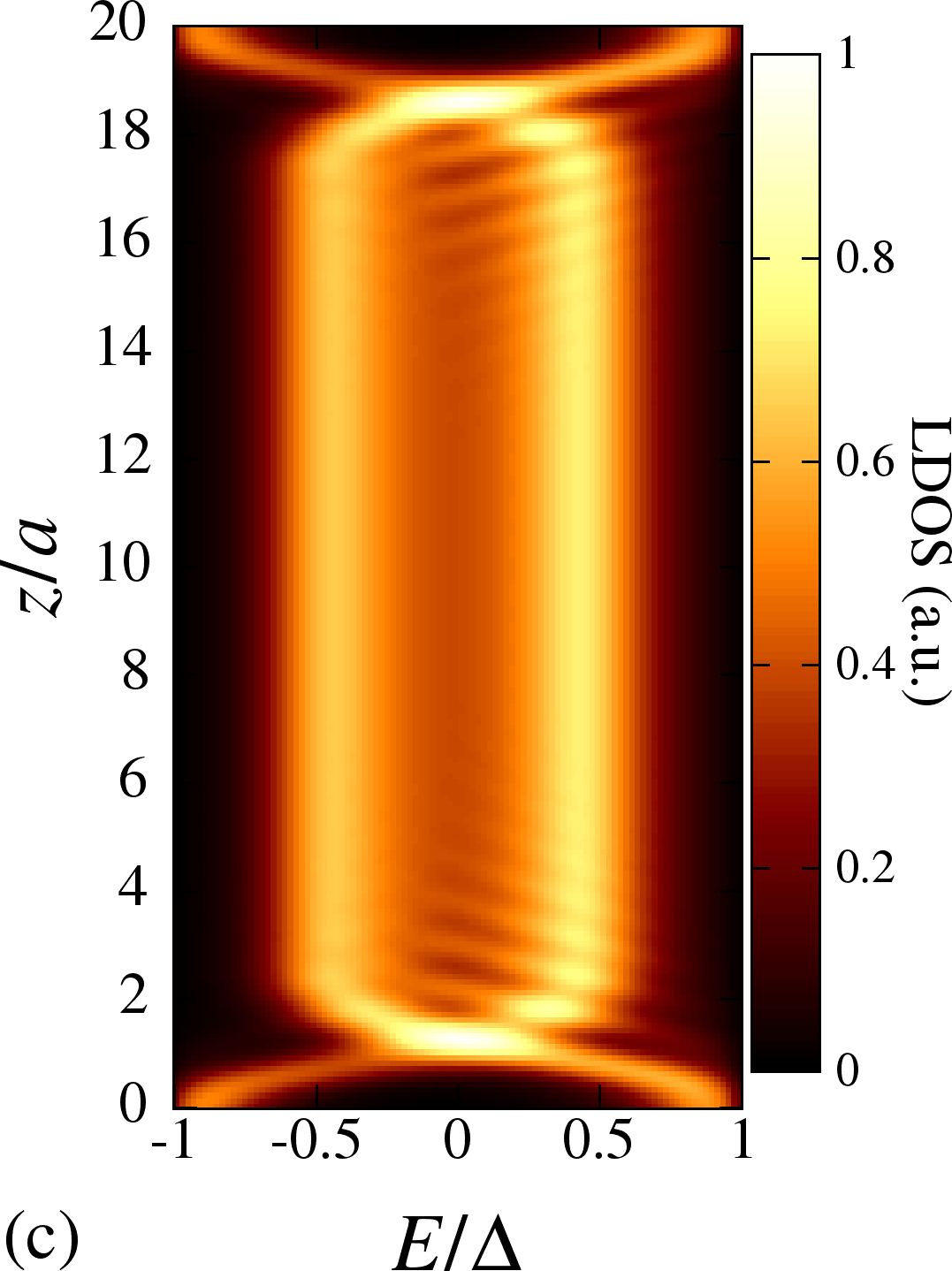}\\[0.1cm]\includegraphics[height=4.7cm]{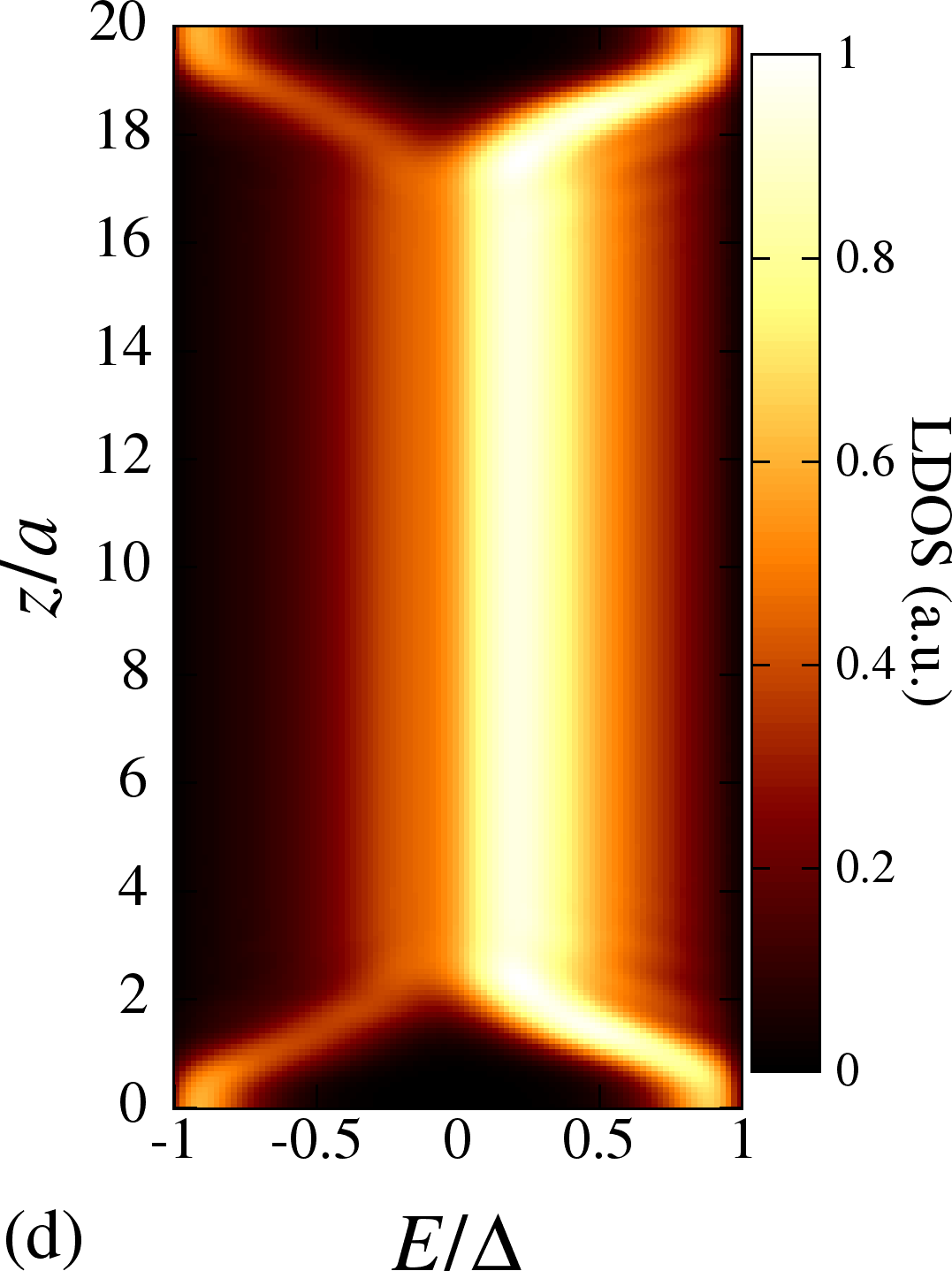}\hspace{0.02\columnwidth}\includegraphics[height=4.7cm]{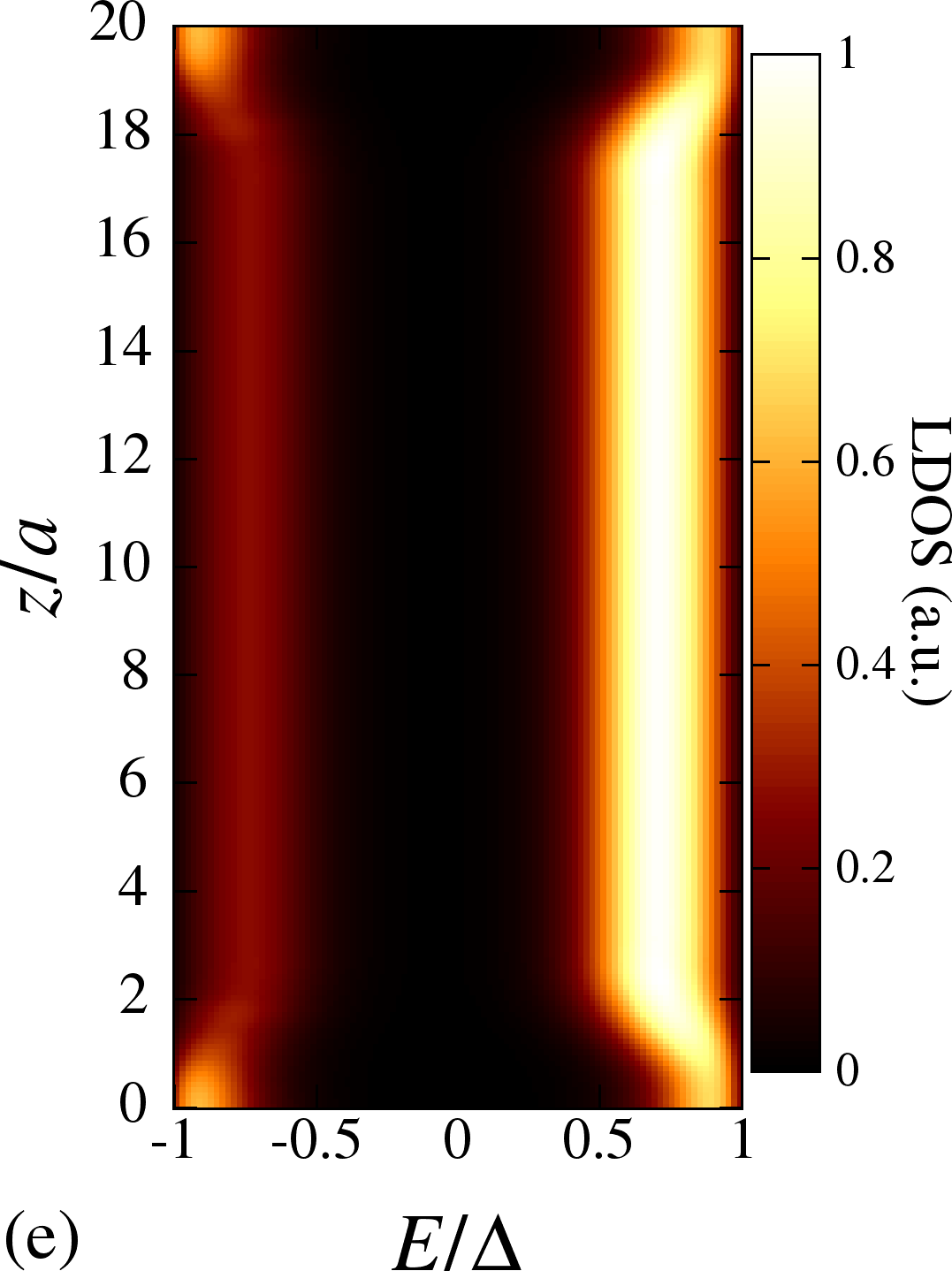}\hspace{0.02\columnwidth}\includegraphics[height=4.7cm]{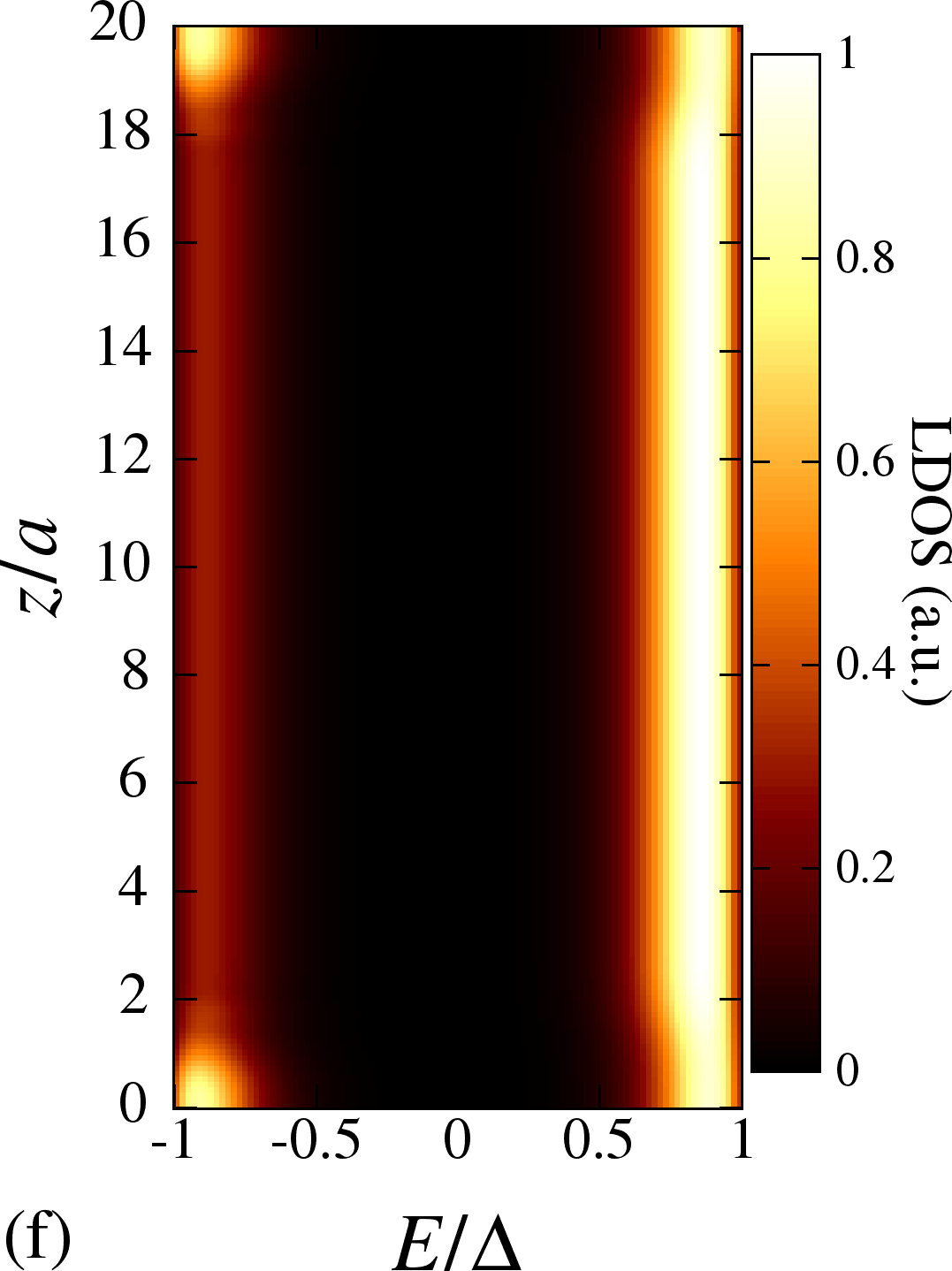}
\par\end{centering}

\caption{Numerical results for the energy-resolved LDOS along the wire
  for the different $m$ channels: (a) $m=0$ [also Fig. (3)(b) of the
    main text]; (b) $m=1$; (c) $m=2$; (d)
  $m=3$; (e) $m=4$; and (f) $m=5$. In each figure the LDOS is
  normalized by its largest value. The potentials in the ferromagnetic
  wire are $V_\uparrow = 0.04\mu$ and $V_\downarrow = 0.6\mu$. 
\label{fig:azimuthal}}
\end{figure}

The subgap LDOS in the wire for the $m=0\ldots5$ YSR states is shown
in~\fig{fig:azimuthal}. As for the $m=0$ results shown in the main text as
Fig.~(3)(b) [and reproduced here in~\fig{fig:azimuthal}(a)], we
observe a significant difference between the states in the bulk and at
the wire ends: the bulk is characterized by broad peaks centered at
the location of the VHSs observed in~\fig{fig:YSRbands}, but within
the rounded ends of the wire these peaks move towards the gap
edges. For the $m\leq2$ states, the $k_z=0$ VHS is located below zero
energy, and so the shifting of the bands leads to an enhancement of
the zero-energy LDOS at the wire ends. The potential is not strong to
cause the $m>2$ YSR bands to cross zero energy, however, and so the
VHSs are not shifted through zero energy at the wire ends and these
bands do not contribute to the zero-energy LDOS
enhancement. Nevertheless, these bands would naturally explain the
subgap features at nonzero energy observed in addition to the
zero-energy peak at the wire ends
in~\cite{Nadj-Perge2014Observation}. However, to quantitatively model
these experiments we would need to be able to weight the contributions
from the different $m$ channels. This requires a detailed
understanding of the coupling to the STM tip which is well beyond the
scope of the current work.


\end{document}